\documentclass[preprint,showpacs,preprintnumbers,amsmath,amssymb,showkeys,aps,prd,a4paper,byrevtex]{revtex4}

\usepackage[dvips]{graphicx}
\usepackage{dcolumn}
\usepackage{bm}
\usepackage{psfig}
\usepackage{amsfonts}
\usepackage{amssymb,amscd}

\begin{document}

\title{Dynamical gluon mass and $pp$ elastic scattering}

\author{M. B. Gay Ducati}
\email{gay@if.ufrgs.br}
\author{W. Sauter}
\email{sauter@if.ufrgs.br}
\affiliation{Instituto de F\'{\i}sica, Universidade Federal do Rio Grande do Sul, Caixa Postal 15051, CEP 91501-970, Porto Alegre, RS, Brasil
}

\date{\today}

\begin{abstract}
The application of a modified gluon propagator and an infrared frozen running coupling constant in the description of the proton-proton elastic scattering is investigated. The model based on the exchange of the Pomeron as a pair of non-perturbative gluons is modified to include the frozen coupling constant in the infrared kinematic region. A good fit to the data of the differential cross section for different center of mass energies (53 GeV and 1.8 TeV) is obtained.
\end{abstract}



\maketitle

\section{Introduction \label{sec:intr}}

The Pomeron has an important role in the elastic scattering, confirmed by the description of the experimental data of the hadronic collisions and diffractive dissociation~\cite{fr97,bp02,ddln02}. In the framework of the Regge theory~\cite{ddln02}, the Pomeron is the exchange of the vacuum quantum numbers, with the scattering amplitude given by
\begin{equation}
\mathcal{A}(s,t) \propto s^{\alpha_{\mathbb P}(t)-1}, \quad\alpha_{\mathbb P}(t) = \alpha(0) + \alpha^\prime(t)
\end{equation}

The most simple picture within Quantum Chromodynamics (QCD) for the Pomeron is the exchange of a pair of gluons, as showed by Low~\cite{lf75} and Nussinov~\cite{ns75}. However, the applicability of perturbative QCD (pQCD) in the description of soft processes is not free of troubles, specially in the singularity structure of the scattering amplitude in the kinematic regime of low momentum transfer. The singularities in this regime include a large value of the strong coupling constant and collinear divergences in the gluon and quark propagators. Usually, when the standard gluon propagator is employed, a divergence (pole) exists when the momentum transfer goes to zero. A solution to this difficulty is the introduction of a cut by hand or the use of the spontaneous symmetry breaking mechanism to generate a mass to the gluon, with the price of the introduction of the Higgs bosons to the theory. Although a bare mass for the gluon (corresponding to a massive propagator) has problems with unitarity and renormalization, this propagator has some interesting phenomenological features~\cite{fpp99}.

Landshoff and Nachtmann (LN)~\cite{ln87} proposed a phenomenological model to overcome these difficulties, arguing that in this kinematic region the non-perturbative effects are important. In this model, these effects change the gluon propagator, which is related with the QCD vacuum through the gluon condensate~\cite{ddln02}.

Ross~\cite{dr89} use the ideas of the LN model in a phenomenological model, in which the Pomeron exchange is described as in the Balitski-Fadin-Kuraev-Lipatov (BFKL)~\cite{fr97} approach, meaning a gluon ladder with the exception of the gluon propagator, modified in the infrared region. The correct model must include the infrared modification of all the Green functions as quark propagator and vertex parts, but the hope is that all infrared contributions can be included in the gluon propagator. These ideas are employed by Cudell and Ross~\cite{cr91} in the description of the $pp$ elastic scattering in low momentum transfer.

In this paper, we propose a modification in the above model. Despite the fact that a running coupling is a next leading order contribution, we will use a dynamical coupling with a frozen (finite) infrared behavior, but related with the gluon propagator as analyzed in previous works~\cite{amn02,amn04,gds03} to describe the process of $pp$ elastic scattering at different energies.

The paper is organized as follows. In the section \ref{sec:mod}, the previous models of the $pp$ elastic scattering are discussed. In section \ref{sec:gpas}, a brief review of the results of gluon propagators and frozen running couplings are presented. The modified phenomenological model and its results are presented and discussed in the section \ref{sec:mdml}. The conclusions are presented in section \ref{sec:conc}.

\section{The model \label{sec:mod}}

The model employed to describe the proton-proton elastic scattering in this work was proposed by Cudell and Ross~\cite{cr91}, based on the two-gluon model to the Pomeron. A remarkable feature of this model is the strong dependence on the wave functions of the involved hadrons, giving a possible source of uncertainty. In the above work, the LN model~\cite{ln87} for the Pomeron is used to avoid the infrared divergences employing a gluon propagator calculated from Dyson-Schwinger equations as shown in the next section.

The hadron-hadron scattering can be derived from the photon-photon scattering~\cite{glf71} with appropriated form factors. With the hadron constituted by valence quarks, the scattering amplitude for two gluon exchange can be written as~\cite{gs77}
\begin{eqnarray}
\mathcal{A}_2^{h_1h_2}=is\alpha_s^2\,n_{1}n_{2}\,\mathcal{C}\;\int\!d\mathbf{k}_ad\mathbf{k}_b\;\delta^{(2)}\left(\mathbf{\Delta}-\mathbf{k}_a-\mathbf{k}_b\right)\mathcal{D}(\mathbf{k}^2_a)\mathcal{D}(\mathbf{k}_b^2)\times \nonumber \\
\left[\mathcal{E}^{h_1}_1(\mathbf{k}_a+\mathbf{k}_b)-\mathcal{E}^{h_1}_2(\mathbf{k}_a,\mathbf{k}_b)\right]\left[\mathcal{E}^{h_2}_1(\mathbf{k}_a+\mathbf{k}_b)-\mathcal{E}^{h_2}_2(\mathbf{k}_a,\mathbf{k}_b)\right], \label{eq:ah1h2}
\end{eqnarray}
where $\mathcal{C}$ is the color factor of the process ($\mathcal{C}=8/9$), $n_i$ is the number of valence quarks ($n_{a,b}=3$) and $h_i$ is the type of hadron ($h_{a,b}=p$). After carrying out the integral, the resulting expression is
\begin{equation}
\mathcal{A}^{pp}_2 = 8is\alpha_s^2\,\int\!d\mathbf{k}\,\mathcal{D}(\mathbf{k}^2)\mathcal{D}\left((\mathbf{\Delta}-\mathbf{k})^2\right)\left[\mathcal{E}_1(\mathbf{\Delta})-\mathcal{E}_2(\mathbf{k},\mathbf{\Delta}-\mathbf{k})\right]^2.
\label{eq:app}
\end{equation}

The form factor $\mathcal{E}_1$ can be identified with the elastic Dirac form factor for the proton,
\begin{equation}
\mathcal{E}_1(t=-\mathbf{\Delta}^2) = \frac{4m_p^2-2.79t}{(4m_p^2-t)\left(1-t/0.71\right)^2},
\label{eq:dff}
\end{equation}
where $m_p$ is the proton mass. The second form factor has a more arbitrary structure, being fixed requiring that infrared divergences disappear. A simple form that agrees with this assumption is
\begin{equation}
\mathcal{E}_2(\mathbf{k}_a,\mathbf{k}_b)= \mathcal{E}_1(\mathbf{k}_a^2+\mathbf{k}_b^2-f\mathbf{k}_a\cdot\mathbf{k}_b),
\label{eq:ivff}
\end{equation}
 where the $f$ parameter varyes depending on the specific hadron. In the proton case, its value is choosen such that the quark wave functions are peeked at $\beta =1/3$. In \cite{cr91}, $f=7$, but using other arguments \cite{cn94} the value of $f=1$ is also advocated.

The scattering amplitude can be rewritten as~\cite{cr91} 
\begin{equation}
\mathcal{A}^{pp}(s,t) = 8is\alpha_s^2\left({\mathcal T}_1-{\mathcal T}_2\right), \label{eq:app2}
\end{equation}
where
\begin{subequations}
\begin{eqnarray}
\mathcal{T}_1 &=& \int\!d^2\mathbf{k}\; \mathcal{D}\left(\frac{\mathbf{q}}{2}+\mathbf{k}\right) \mathcal{D}\left(\frac{\mathbf{q}}{2}-\mathbf{k}\right) G_p^2(q,0), \label{e:sct1} \\
\mathcal{T}_2 &=& \int\!d^2\mathbf{k}\; \mathcal{D}\left(\frac{\mathbf{q}}{2}+\mathbf{k}\right) \mathcal{D}\left(\frac{\mathbf{q}}{2}-\mathbf{k}\right) G_p\left(q,k-\frac{q}{2}\right) \left[ 2G_p(q,0) - G_p\left(q,k-\frac{q}{2} \right)\right], \label{e:sct2}
\end{eqnarray}
\end{subequations}
where $s,t=-q^2$ are the Mandelstam variables and $G_p(q,k)$ is a convolution of the proton wave functions,
\[ G_p(q,k) = \int\!d^2p\,d\alpha\;\psi^\ast(\alpha,p)\;\psi(\alpha,p-k-\alpha q), \]
and can be related with the Dirac form factor as follows,
\begin{subequations}
\begin{eqnarray}
G_p(q,0) &=&  F_1(q^2) \label{e:ff1} \\
G_p\left(q,k-\frac{q}{2}\right) &=& F_1\left(q^2+9\left|k^2-\frac{q^2}{4}\right|\right)
\end{eqnarray}
\end{subequations}

The terms in eqs.~(\ref{e:sct1},\ref{e:sct2}) have a simple interpretation: the former comes from Feynman diagrams where the gluons are connected to the same quark in the proton. The last one comes from diagrams where the gluons are connected to different quarks in the proton~\cite{cr91}.

The total cross section is given by the optical theorem, $\sigma_{\rm tot}^0 = \mathcal{A}_2^{pp}(s,0)/is$, as well as the differential cross section $d\sigma^0/dt = |\mathcal{A}_2^{pp}(s,t)|^2/16\pi s^2$. For low momentum transfer, the elastic differential cross section is fitted by a exponential expression, $d\sigma/dt = A\,e^{Bt} = \sigma_{\rm tot}^2\,e^{Bt}/(16\pi)$, where $B$ is the logarithmic slope, given by $B=d/dt[\ln(d\sigma/dt)]|_{t=0}$. 

With the scattering amplitude above, a finite result for the total cross section is obtained due to the cancelation of the infrared divergences~\cite{cr91,hkn93}. Otherwise, the differential cross section has an unphysical result ($B(t=0)\rightarrow\infty$) if the perturbative gluon propagator is employed. When the Landshoff-Nachtmann model is used, the contribution from the term ${\mathcal T}_2$ is negligible and the remaining infrared divergences are regularized by the modified gluon propagators.

However, the two gluon description of the scattering amplitude is a crude approximation, which does not yield any energy dependence in the above cross sections. The reason is the absence of the gluon ladders in the $s$ channel, as in the BFKL framework, which gives an energy dependence in the cross sections. To restore this dependence, we assume that the amplitude above only describes the energy independent part of the cross section and an extra term, introduced {\it ad hoc} gives the Regge energy behavior, namely,
\[ \mathcal{A}^{pp}(s,t) \rightarrow \left(\frac{s}{s_0}\right)^{\alpha_{\mathbb P}(t)-1}\mathcal{A}^{pp}(s,t), \]
where $\alpha_{\mathbb P}(t)=\alpha(0) + \alpha^\prime(t)$ is the (soft) Pomeron trajectory (with $\alpha(0)=1.08$ and $\alpha^\prime(t) \simeq 0.25$ GeV$^{-2}$~\cite{dl92}) and $s_0$ is an energy scale. The extra Regge term gives the following cross sections,
\begin{subequations}
\begin{eqnarray}
\sigma_{\rm tot} &=& \left(\frac{s}{s_0}\right)^{\alpha_{\mathbb P}(0)-1}\sigma_{\rm tot}^0, \label{e:tostf} \\
\frac{d\sigma}{dt} &=& \left(\frac{s}{s_0}\right)^{2\alpha_{\mathbb P}(t)-2}\frac{d\sigma^0}{dt} \label{e:todsf}.
\end{eqnarray}
\end{subequations}

The above model was used by \cite{cr91,hkn93,hpr96} where the non-perturbative propagator comes from different methods, discussed in the next session. The experimental data are ISR results for elastic proton-proton scattering at $\sqrt{s}=$ 53 GeV~\cite{isr84}.

\section{Gluon propagator and running coupling constant \label{sec:gpas}}

A central idea of the LN model for the Pomeron is the infrared modified gluon propagator, where the modifications are induced by the QCD vacuum effects, and its non perturbative character required the application of methods of the same nature to obtain the gluon propagator. The most popular methods are the Dyson-Schwinger equations (DSE)~\cite{rw94,as01} and the numerical computational simulations on lattice field theory~\cite{2roth}.

The DSE's are an infinite system of non-linear, coupled integral equations which relate the different Green functions of a quantum field theory. In the case of QCD, due to the complexity of the system to be solved, there are several analytical approximations in the literature to obtain a solution. A discussion among the different methods and approximations employed can be found in \cite{rw94,as01}.

The lattice field theory, in opposition to the DSE, is a powerful pure numerical method, where the main idea is that the divergences of the theory are regularized by the discretization of the space-time. Otherwise, the lattice approximation is not free of problems: finite size of the lattice, spacing between the sites of the lattice and fermion simulation. 

Due to the approximations employed in the resolution of the problems pointed above, there are many different solutions for both methods in the literature. We focus here on the solutions which have the particular property of a dynamical gluon mass. A gluon propagator with this feature was used in successful descriptions of several processes~\cite{amn02,amn04,gds03}. 
The first solution for the gluon propagator which will be used in this work is the Cornwall's solution~\cite{jmc82}, from DSE's in the axial gauge with the use of a technique of resummation of Feynman diagrams,
\begin{equation}
D^{-1}_{C}({\mathbf q}^2) = \left[ {\mathbf q}^2 + m^2_C({\mathbf q}^2) \right] b g^2 \ln \left( \frac{{\mathbf q}^2 + 4 m^2_C({\mathbf q}^2)}{\Lambda^2_{\rm QCD}} \right), \label{e:dgco}
\end{equation}
with 
\begin{equation}
m^2_C({\mathbf q}^2) = m^2_g \left[ \ln\left(\frac{{\mathbf q}^2 + 4m^2_g}{\Lambda^2_{\rm QCD}}\right)\left/ \ln\left( \frac{4m^2_g}{\Lambda^2_{\rm QCD}} \right)\right. \right]^{-12/11},
\end{equation}
where $m_g=$ 500 $\pm$ 200 MeV for $\Lambda_{\rm QCD}$ = 300 MeV and $b=33/(48\pi^2)$ is the leading order coefficient of the $\beta$ function. A variation of the above propagator is given by
\begin{equation}
D_{C^\prime}({\mathbf q}^2) = \frac{1}{{\mathbf q}^2 + m^2_C({\mathbf q}^2)}. \label{e:dgcv}
\end{equation}
Other solution is obtained by H\"abel {\it et al.}~\cite{hab90,hab90a} using another type of approximation, which gives the following propagator,
\begin{equation}
D_{H}({\mathbf q}^2) = \frac{{\mathbf q}^2} {{\mathbf q}^2+b^4}, \label{e:dgh}
\end{equation}
where $b$ is a free parameter. Gorbar and Natale~\cite{gn00} use the propagator
\begin{equation}
\left[D_{GN}({\mathbf q}^2)\right]^{-1} = {\mathbf q}^2 + \mu^2_g \Theta(\xi^\prime\mu^2_g-{\mathbf q}^2) + \frac{\mu^4_g}{{\mathbf q}^2}\Theta({\mathbf q}^2-\xi^\prime\mu^2_g) \label{e:dggn}
\end{equation}
to calculate the vacuum QCD energy through effective potentials, where $\Theta$ is the step function, $\mu_g$ is related with the gluon condensate and $\xi^\prime$ is a calculated parameter. An alternative form used in this work is 
\begin{equation}
\left[D^\prime_{GN}({\mathbf q}^2)\right]^{-1} = {\mathbf q}^2+m^2_{GN}\Theta(m^2_{GN}-{\mathbf q}^2) + \frac{m^4_{GN}}{{\mathbf q}^2}\Theta({\mathbf q}^2-m^2_{GN}) \label{e:dggnv},
\end{equation}
where $m_{GN}$ is a mass parameter. A recent result for the gluon propagator, obtained from the DSE's in the Mandelstam approximation, is due to Aguilar and Natale~\cite{an04a}, where the gluon propagator is well fitted by the expression
\begin{equation}
D_{AN}({\mathbf q}^2) = \frac{1}{{\mathbf q}^2+\mathcal{M}^2(\mathbf{q}^2)},\quad\mathcal{M}^2(\mathbf{q}^2)=\frac{m_0^4}{\mathbf{q}^2+m_0^2}. \label{e:dgan}
\end{equation}

Another recent result is the solution of DSE's in the Landau gauge calculated by Alkofer and collaborations~\cite{as01,sha97,fa03,adfm04}, where the gluon propagator are given by \cite{adfm04}
\begin{equation}
D_{AL}({\mathbf q}^2) = \frac{\omega}{{\mathbf q}^2} \left[\frac{{\mathbf q}^2}{\Lambda^2_{\rm QCD}+{\mathbf q}^2}\right]^{2\kappa}\left(\alpha_s^{\rm (AL)}({\mathbf q}^2)\right)^{-\gamma},
\label{e:dgal}
\end{equation} 
where $\omega=2.5$, $\Lambda_{\rm QCD} = 510$ MeV, $\kappa \approx 0.595$, $\gamma = -13/22$ and $\alpha_s^{(\rm AL)}({\mathbf q}^2)$ is the running coupling constant,
\begin{equation}
\alpha_s^{(\rm AL)}({\mathbf q}^2) = \frac{1}{1+({\mathbf q}^2/\Lambda^2_{\rm QCD})} \left[ \alpha_s(0) + \frac{4\pi}{\beta_0} \frac{{\mathbf q}^2}{\Lambda^2_{\rm QCD}} \left( \frac{1}{\ln({\mathbf q}^2/\Lambda^2_{\rm QCD})} + \frac{1}{1-({\mathbf q}^2/\Lambda^2_{\rm QCD})} \right)  \right],
\label{e:arcc}
\end{equation}
where $\alpha_s(0) \approx 2.972$ and $\beta_0 = 11$.

\begin{figure}[t]
\begin{center}
\begin{tabular}{cc}
\scalebox{0.5}{\includegraphics*[25,35][510,520]{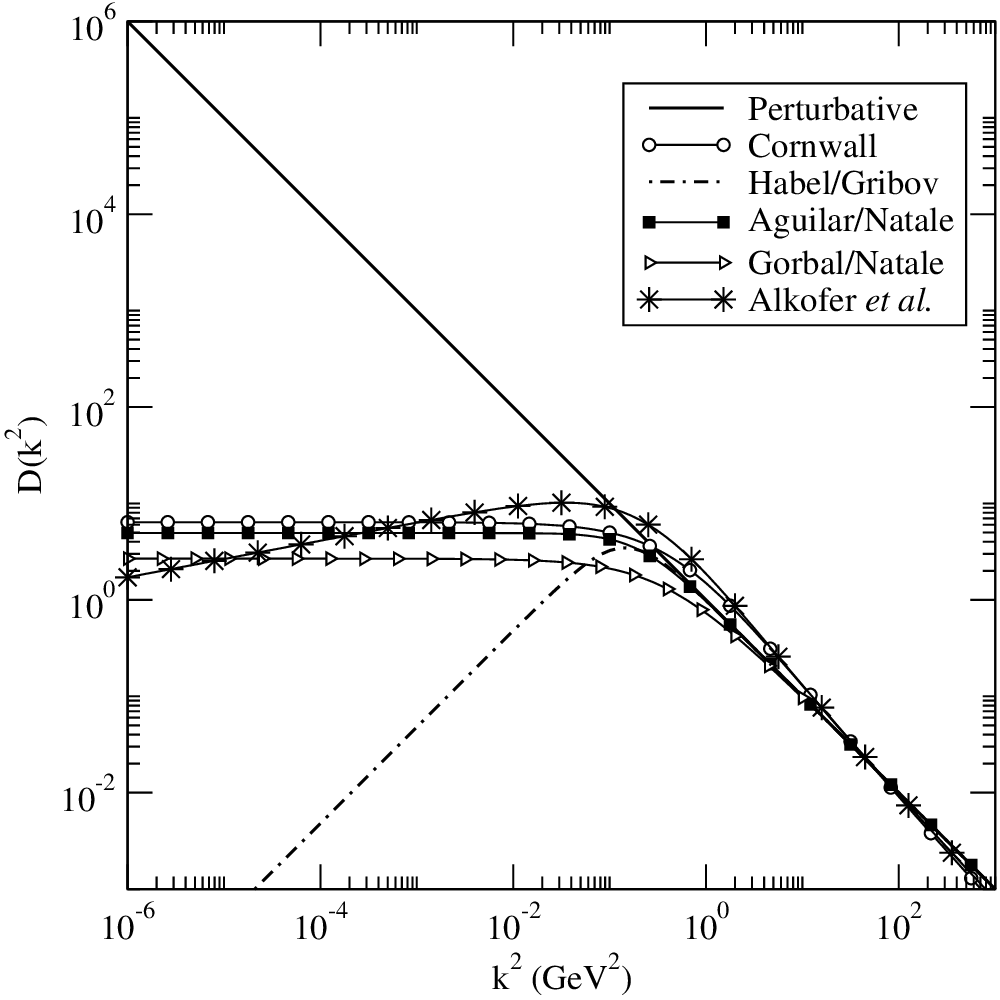}} &
\scalebox{0.5}{\includegraphics*[25,35][510,520]{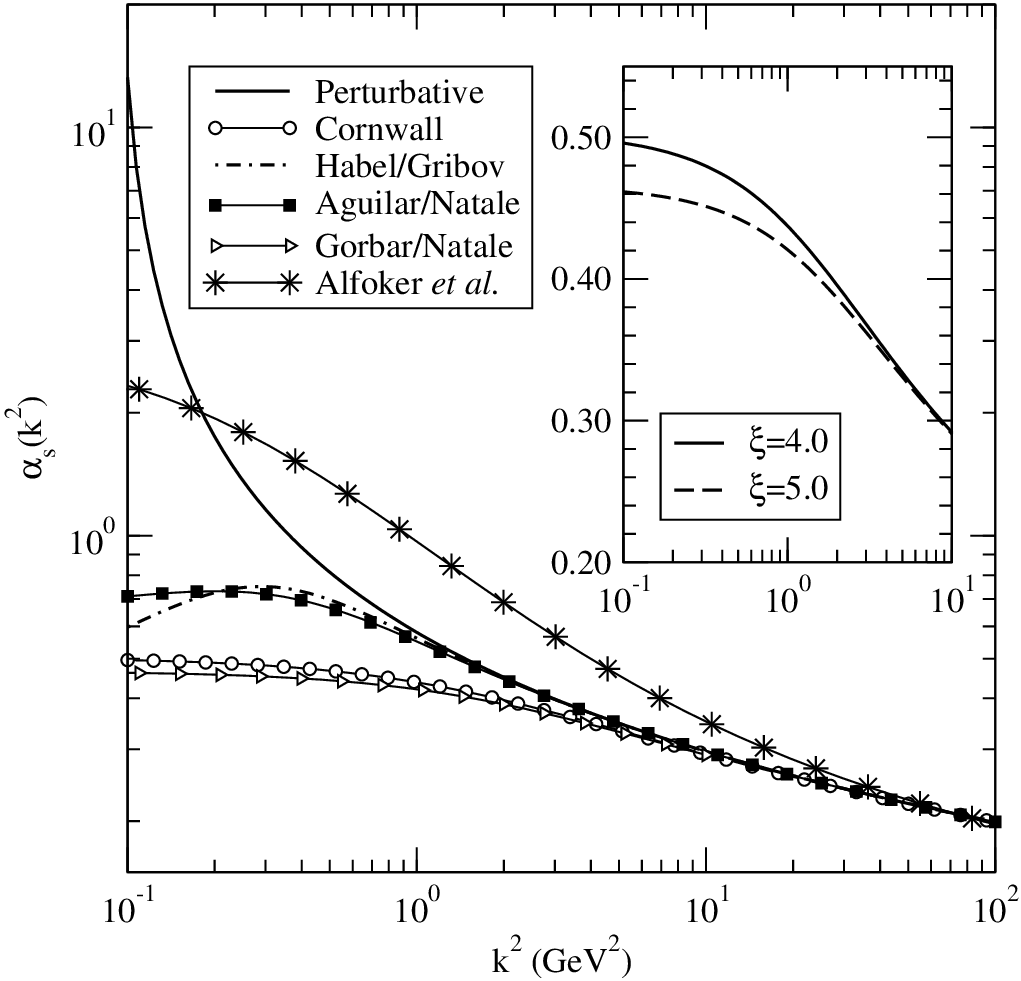}} \\
(a) & (b)
\end{tabular}
\end{center}
\caption{(a) Distinct modified gluon propagators compared with the perturbative one. The propagators are indicated in the plot legend. (b) Comparison between different results for the running coupling constant. In the detail, two results of the Cornwall frozen coupling with two different $\xi$'s. The parameters of the curves are given in the text.}  \label{f:dg}
\end{figure}

The momentum dependences of the gluon propagators are displayed in the figure \ref{f:dg}, in comparison with the perturbative one. The most remarkable feature is the infrared behavior ($\mathbf{k}\rightarrow 0$), where exists a finite value in the case of the Cornwall propagator, for example; a zero value, as in the H\"abel/Gribov propagator, and the divergent perturbative propagator. Another remarkable fact is the same ultraviolet behavior ($\mathbf{k}\rightarrow \infty$) of the modified propagators and the perturbative one.

Nevertheless, there is a narrow relation between the gluon propagator and the running coupling, widely discussed in \cite{amn02,amn04}. The phenomenological applications of a frozen coupling constant include heavy quarkonia decays~\cite{mn00}, meson form factors~\cite{amn02,gds03} and other observables~\cite{amn04}. In some of the above results (specially in \cite{gds03}), the data available are described when the modified gluon propagator and the frozen coupling constant are used at the same time.

The analytical expression for the running coupling is obtained following the Cornwall's solution~\cite{jmc82,cp91}, namely,
\begin{equation}
\alpha_s(k^2)=4\pi\left/ \beta_0\ln\left(\frac{k^2+\xi m_{\mathcal P}^2(k^2)}{\Lambda^2_{\rm QCD}}\right) \right. ,
\label{e:frcc}
\end{equation}
where the explicit momentum dependence is the same as the one loop perturbative calculation, but with an extra term, $\xi m_{\mathcal P}^2(k^2)$, resulting in the frozen infrared behavior where $\xi$ is a parameter that varies with the propagator and measures the frozeness of the coupling. In the original work~\cite{jmc82,cp91}, the value of this parameter is obtained as $\xi\approx 4$ or larger.  The massive term, $m_{\mathcal P}(k^2)$, also depends on the gluon propagator employed, being the dynamical mass term of the propagator. For example, in the case of the Aguilar and Natale propagator, eq. (\ref{e:dgan}), this term is $m_{\mathcal P}(k^2)=\mathcal{M}^2(k^2)$. The exception is the Alkofer case, with a proper expression for the running coupling constant, displayed above.

In the fig.~\ref{f:dg}, we present the different $\alpha_s$'s (given by eq.(\ref{e:frcc})), where the freezing of the running coupling constant, the matching of the frozen running coupling, and the behavior of the perturbative result in high momentum are very clear. The last result comes from the behavior of the massive term ($m_{\mathcal P}(k^2)$) which goes to zero when the momentum is large in all the propagators employed in this work (in the case of massive propagator, the mass is small). The parameters of some propagators ($m_g$, $b$, $m_0$,\ldots) are determined when the results of the elastic $pp$ scattering are fitted (see below) with exception of the Alkofer propagator. The figure also shows the $\xi$ parameter dependence of the Cornwall solution, eq. (\ref{e:dgco}). As seeing in fig.~\ref{f:dg} and also in the results of the model, there is a small difference for distinct $\xi$ values.

\section{The modified model \label{sec:mdml}}

In the previous model for the scattering amplitude, the running coupling constant was consider fixed. In \cite{gdhn93}, the value of the coupling was determined throught the effective Pomeron coupling with the hadrons, which depends on the gluon propagator and has well defined experimental value.

In this work, motivated by the good results of the previous calculations with a gluon propagator and a frozen running coupling~\cite{gds03}, we modify the previous model of $pp$ scattering, considering now a frozen running coupling constant and a modified gluon propagator. We emphasize that this is a phenomenological modification, since a running coupling is a next leading order contribution and the model employed is a first order contribution to the Pomeron exchange. As usual in the previous works, the scale of the running coupling constant is chosen as the incoming momentum into the quark-gluon vertex.

The scattering amplitude in the modified model then reads
\begin{eqnarray}
\mathcal{A}^{pp}_2(s,t) &=& 8is\left(\frac{s}{s_0}\right)^{\alpha_{\mathbb P}(t)-1}\int\!d^2\mathbf{k}\; \alpha_s\left(\frac{\mathbf{q}}{2}+\mathbf{k}\right) \mathcal{D}\left(\frac{\mathbf{q}}{2}+\mathbf{k}\right) \alpha_s\left(\frac{\mathbf{q}}{2}-\mathbf{k}\right) \mathcal{D}\left(\frac{\mathbf{q}}{2}-\mathbf{k}\right) \times \nonumber \\
& &\left[ G_p(q,0) - G_p\left(q,k-\frac{q}{2}\right) \right]^2. \label{e:appm}
\end{eqnarray} 
where the terms from eqs. (\ref{e:sct1}) and (\ref{e:sct2}) were combined as one term only and the running coupling terms are included in the integrand. 

\begin{figure}[t]
\begin{center}
\begin{tabular}{cc}
\scalebox{0.5}{\includegraphics*[70,35][535,535]{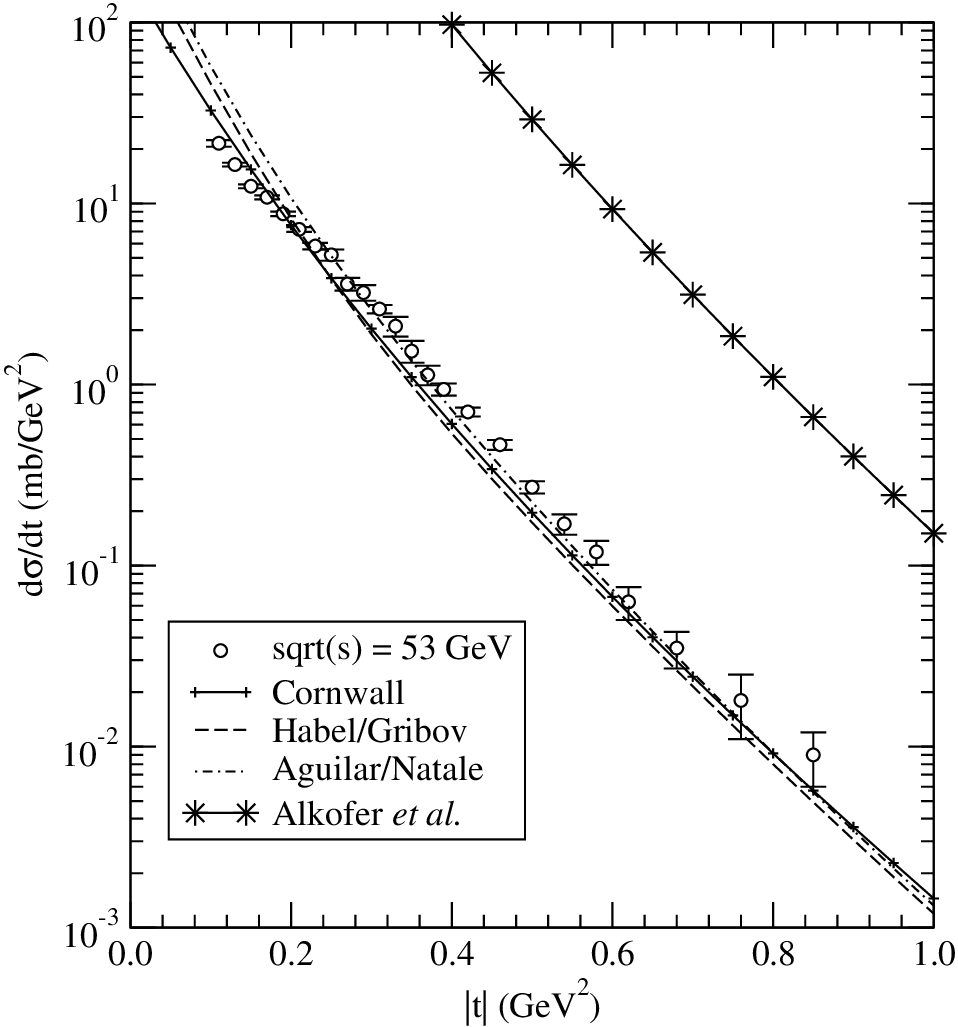}} & 
\scalebox{0.5}{\includegraphics*[70,35][535,535]{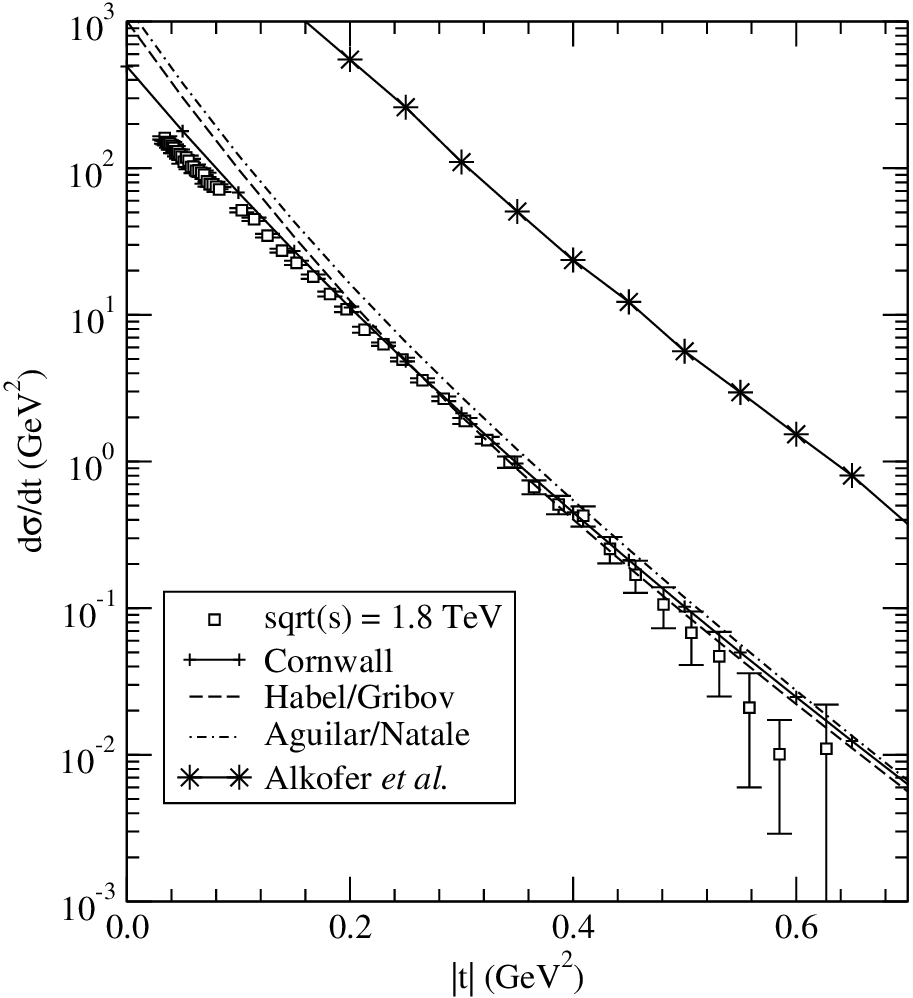}}
\end{tabular}
\end{center}
\caption{Results of modified model, fitted to the $pp$ elastic scattering data with $\sqrt{s}=$ 53 GeV\protect{~\cite{isr84}}(right) and $\sqrt{s}=$ 1.8 TeV\protect{~\cite{e71090}}(left) with different gluon propagator and frozen coupling constants. The parameters of the fits are given in the text.}  
\label{f:dsdt}
\end{figure}

The differential cross section is then calculated using the eq. (\ref{e:appm}) into eq. (\ref{e:todsf}), varying the different propagators as well as the running couplings constants, as pointed out above. The results of the calculation are displayed in the fig.~\ref{f:dsdt} for two sets of experimental data from different center of mass energies: $\sqrt{s}=$ 53 GeV from ISR~\cite{isr84}, and $\sqrt{s}=$ 1.8 TeV from Tevatron~\cite{e71090}.

We obtain a good agreement with the data in both energies and in the displayed range of transfered momentum. The slope of the curves are in accordance with  data, due the {\it ad hoc} term in the amplitude. The energy independent part of the amplitude, modeled by the eq.~(\ref{eq:app2}) gives a contribution on the $t$ behavior and the intercept ($d\sigma/dt|_{t=0}$) of the curve.

Some considerations about the results can be made. First, the variation of the Cornwall propagator, eq. (\ref{e:dgcv}), and the Gorbar/Natale one, eq. (\ref{e:dggnv}) are not displayed in the fig. \ref{f:dsdt} since these propagators give results quite similar to the presented ones.

The values of the gluon propagator parameters used to obtain the results shown (or not) in the figure are: Cornwall, $m_g=0.53$ GeV (with $\Lambda_{\mathrm QCD}=$ 0.3); Cornwall variant, $m_g=$ 0.3 GeV; Gorbar/Natale, $m_g=$ 0.4; H\"abel/Gribov, $b=$ 0.3 and Aguilar/Natale, $m_0=$ 0.35. The calculation is sensitive to the value of the parameters, giving larger modifications when the massive parameter is changed. In opposition, when the $\xi$ parameter is changed, the change in the result is significantly smaller than with the mass parameter. The reason is the small modification in the running coupling when $\xi$ value is changed, as pointed out in the previous section. Thus, we maintain the value of the $\xi$ fixed in 4. The results of the modified model show that the mass parameter, for this process, is in the range of 300 MeV to 600 MeV, in agreement with the previous works which employ the modified gluon propagators.

When the Alkofer {\it et al.} propagator, eq. (\ref{e:dgal}), and the corresponding running coupling constant, eq. (\ref{e:arcc}), with the original set of parameters is employed, we obtain a result that overestimates the experimental data in both energies. Similar result was found by \cite{amn02,amn04} for another observables.

\section{Conclusions \label{sec:conc}} 

In this work, we consider a frozen running coupling instead of a fixed one in a model for the amplitude of $pp$ elastic scattering with the exchange of two non-perturbative gluons. The same idea was employed in another processes, which a small momentum transfer as, for example, the description of meson form factors~\cite{amn02,gds03}, with successful description of the data. 

In the present process, we obtain a quite good agreement with experimental data for distinct energies. The results obtained are along in the same lines of the previous attempts to describe the process, but considering other propagators. The results show that the employ of modified gluon propagators is valuable in phenomenology, but with some caution, since the model used to construct the scattering amplitude is a first approximation. A possible extension is to employ a complete set of Green functions: the quark and gluon propagator and the quark-gluon vertex.

The application of the distinct results for the gluon propagator (and Green functions) found in the literature is also a field to test these solutions as well as the frozen running coupling constant. In the present work, the results point out to a gluon propagator with dynamical mass and frozen running coupling constant, although the present results in addition with the previous ones~\cite{hkn93,amn02,amn04,gds03}, do not determine which is the most suitable expression for the propagator, mainly due the approximations employed and the still low knowledge in the literature on the interplay of whole aspects involved in these calculations. The consistency of these ideas used here can be checked in another type of diffractive processes, for example, diffractive photo-production of light vector mesons, which should deserve some analysis in this framework. 

\begin{acknowledgments}
This research was supported by the Conselho Nacional de Desenvolvimento Cient\'{\i}fico e Tecnol\'ogico (CNPq). 
\end{acknowledgments}

\bibliography{bibdefs,bfkl,difracao,eds,livros,ln,mesvec,outros,pp,qcd,rede,decmff}
\bibliographystyle{unsrt}

\end{document}